\documentclass[conference]{IEEEtran}
\IEEEoverridecommandlockouts
\usepackage{cite}
\usepackage{amsmath,amssymb,amsfonts}
\usepackage{algpseudocode}
\usepackage{algorithm}
\usepackage{graphicx}
\usepackage{textcomp}
\usepackage{xcolor}

\def\BibTeX{{\rm B\kern-.05em{\sc i\kern-.025em b}\kern-.08em
    T\kern-.1667em\lower.7ex\hbox{E}\kern-.125emX}}
\begin{document}

\title{GGArray: A Dynamically Growable GPU Array\\
\thanks{This research was supported by the Temporal research group, the ANID Fondecyt grant \#1221357 and the Patagón supercomputer of Universidad Austral de Chile (FONDEQUIP \#EQM180042).}
}

\author{\IEEEauthorblockN{Enzo Meneses}
\IEEEauthorblockA{\textit{Instituto de Informática} \\
\textit{Universidad Austral de Chile}\\
Valdivia, Chile \\
enzo.meneses@alumnos.uach.cl}
\and
\IEEEauthorblockN{Cristóbal A. Navarro}
\IEEEauthorblockA{\textit{Instituto de Informática} \\
\textit{Universidad Austral de Chile}\\
Valdivia, Chile \\
cristobal.navarro@uach.cl}
\and
\IEEEauthorblockN{H\'ector Ferrada}
\IEEEauthorblockA{\textit{Instituto de Informática} \\
\textit{Universidad Austral de Chile}\\
Valdivia, Chile \\
hferrada@inf.uach.cl}
}

\maketitle

\begin{abstract}
This work presents a dynamically growable GPU array (GGArray) fully implemented in GPU that facilitates the programming of GPU applications with dynamic memory. The GGArray is based on an array of LFVectors, taking advantage of the GPU architecture and its synchronization at block level. The GGArray is compared to other state of the art approaches such as a pre-allocated static array and a semi-static array that needs to be resized through communication with the host. Experimental evaluation shows that the GGArray achieves an efficient memory usage close to the optimal and not greater than $2\times$ the needed memory, as well as a competitive insertion/resize performance, but it is slower for regular parallel memory accesses. Given these results, the GGArray is a potentially useful structure for applications with high uncertainty on the memory usage as it does not require pre-allocating GPU VRAM for the worst case scenario. It can also be useful in applications that exhibit phases in terms of memory behavior, such as an insertion phase followed by a regular r/w GPU phase. In these cases, the GGArray can be used for the first phase and then data can be flattened for the second phase in order to allow the regular and faster GPU memory accesses to take place. These results constitute a step towards achieving a parallel efficient C++ like vector for modern GPU architectures.  
\end{abstract}


\section{Introduction}
GPUs have become a great contribution in HPC, scientific simulations and other applications because of their high parallel performance and energy efficiency \cite{Navarro2014ASO}. Furthermore, recent GPU improvements such as tensor cores and ray tracing cores have cemented their use in certain areas that receive an even greater benefit from these technologies. GPUs are especially useful when dealing with grid-like structured data such as arrays or matrices, offering up to an order of magnitude of speedup over a CPU based approach. It is also known that when dealing with graphs, sparse matrices and other irregular structures that do not follow a static grid-like layout, the GPU speedup is not as high because of several memory access performance penalties \cite{cudaBestPractices}.

This problematic also extends to the use of dynamic memory. Given the impossibility to maintain data contiguously in memory when dynamically allocating it without any kind of global synchronization, it is natural that the use of dynamic memory does not provide the same speedups as the ones obtained with static structured data. Although there are several studies on graph algorithms and sparse matrices \cite{Busato2018HornetAE,Awad2020DynamicGO,King2016DynamicSA,Sha2017AcceleratingDG} for GPUs, some of which also explore the use of dynamic memory, there has not been an attempt to implement a more general dynamic array that works on GPU. Dynamic arrays, such as for example C++ vectors,  are one of the most commonly used structures in programming languages and some of them do not even include static arrays in their standard library (e.g. python). Dynamic arrays provide an easier way of programming and allow simpler code designs due to its capacity to grow or shrink as required during execution.

This simplicity is a valuable feature for a large community who's work is more focused on developing an application than a tool, i.e., their effort should not be focused on writing the needed data structures. A significant part of the scientific community has these requirements as they focus on the study of certain phenomena that requires the use of intensive computer simulations accelerated by GPU libraries. Currently, it is highly difficult to take advantage of the benefits of dynamic arrays on GPU, and when it is absolutely necessary, it requires a significant amount of low-level programming effort. Also, in many cases this programming effort results in a handmade structure for an specific application, to some degree, making it unusable for other applications. Having a general purpose dynamic array for GPUs would improve the programming model substantially, however, accomplishing an efficient one for GPUs is considered a difficult challenge \cite{8855701}.

One of the first aspects to consider in dynamic arrays is that they are by design slower than static ones due to the work required to maintain data integrity after each operation, especially on GPUs where it is necessary to deal with thousands of parallel operations. In exchange, it offers a more efficient use of VRAM memory, due to its capacity to adjust its size to the amount of data contained at each moment of the execution. This efficient memory usage allows to run more applications simultaneously in a GPU, via concurrent kernel execution, as long as the peak memory consumption doesn't occur at the same time. Much of the knowledge in parallel dynamic arrays has come from the LFVector \cite{Dechev2006LockFreeDR} structure proposed in 2006, which is one of the first works that describes an implementation of a parallel dynamic array for CPUs.    

This work proposes a dynamically growable array for the GPU, named GGArray, which is based on the LFVector idea adapted to the massive parallelism programming model. The structure is divided into blocks, exploiting the asynchronous advantage of GPU thread blocks and diminishing global synchronization issues, however at the cost of having a slower access to its elements. An experimental evaluation is performed in terms of performance and memory usage, showing competitive insertion performance and a more efficient memory usage when compared to other approaches such as static and semi-static arrays. The GGArray is a first step into the construction of a data structure with an interface similar to the C++ vector that works on a massively parallel architecture. 

The remaining sections cover related work (Section \ref{sec:related-work}), current approaches for array data structures and insertion schemes (Section \ref{sec:current-approaches}), the presentation of the proposed GGArray (Section \ref{sec:ggarray}), a theoretical analysis of memory usage (Section \ref{sec:theo-mem-usage}), an experimental evaluation (Section \ref{sec:exp-eval}) and conclusions (Section \ref{sec:conclusions}).

\section{Related Work}
\label{sec:related-work}
While there is no openly available general dynamic array fully implemented in GPU, there are implementations of resizable GPU arrays from the host and a significant amount of research on parallel arrays for multi-core CPUs. Also, there are works that implement hand-tailored GPU-based dynamic memory management for specific scientific applications.

\subsection{GPU Resizable Arrays}
The closest data structures to a dynamic array on a GPU device are offered by the libraries \textit{Thrust}\cite{BELL2012359} and \textit{stdgpu}\cite{Stotko2019stdgpuES}, but none of these manage the dynamic operations fully in GPU. \textit{Thrust} is a well known CUDA library that implements useful data structures to simplify CUDA programming. Among these structures are \textit{host\_vector} and \textit{device\_vector}, dynamic arrays that reside on the host and device memory respectively. But \textit{device\_vector} works like a doubling array and its resizing methods can only be called from the host. On the other hand, there is \textit{stdgpu}, that implements data structures from the C++ STL in CUDA. In this case the vector implementation allows operations to be called from the device, \textit{push\_back} being one of them, but they are implemented with locks, penalizing the potential benefits from parallelization.

\subsection{General Parallel Dynamic Array}
Lock-Free Vector (LFVector) \cite{Dechev2006LockFreeDR} was the first proposed parallel dynamic array and the catalyst for further research on them. It proposes an idea similar to doubling arrays by duplicating the size each time more memory is needed. Differently from doubling arrays it abandons the idea of storing an array contiguously, and divides it into buckets. This difference is important as it avoids the necessity of moving the elements to the new array when resizing, and doesn't require as much synchronization between the distinct threads.

Further research on the topic include improvements to the LFVectors, such as the Wait-Free Vector \cite{Feldman2016AnEW} and new approaches like RCUArray \cite{Jenkins2018RCUArrayAR} which uses the Read-Copy-Update mechanism. From the point of view of massive GPU parallelism, one drawback of the mentioned arrays is that some of their stages rely on synchronization mechanisms thought for CPU architectures, not for GPU ones.

\subsection{GPU Synchronization}
Global synchronization is usually avoided in the GPU, because of the overhead that it introduces. Unfortunately, in some cases it is highly difficult to avoid it. The simplest way to synchronize all threads is by dividing a parallel application into several kernels and using the host as a synchronization barrier. This design makes any data update operation to travel between the host and device in both directions, which can become a performance bottleneck. Because of this, efficient GPU implementations should try to find designs that make synchronization occur inside the device, even if it requires doing it at block-level and not globally.

Research on synchronization includes Fast Barrier Synchronization \cite{Xiao2010InterblockGC} and methods proposed for memory allocation \cite{Gelado2019ThroughputorientedGM} among others. The first work proposes two algorithms for inter-block synchronization. A lock-based method with the use of atomic operations and a lock-free one, which dedicates one block of threads and global memory to indicate whether threads from other blocks are allowed to pass the barrier.
The second work focuses on memory allocation, which they separate into two stages. In the first stage, accounting the available resources, global synchronization is needed, for which they implement semaphores that allow concurrency in the critical section diminishing the principal bottleneck of semaphores. The work also highlights the importance of global synchronization when dealing with dynamic memory or memory allocators.

\subsection{GPU Memory Allocators}
Winter et al. (2021) \cite{Winter2021AreDM} compared and evaluated various memory allocators for NVIDIA GPUs including the allocator provided by the CUDA-Toolkit and non-proprietary allocators starting from XMalloc \cite{Huang2010XMallocAS} and ScatterAlloc \cite{8855701} up to Ouroboros \cite{Winter2020OuroborosVQ}, one of the latest. Although this work does not focus on proposing a new memory allocator, they are relevant as potential tools that can complement and improve the ideas of this work. 

\subsection{Specific GPU Dynamic Applications}
When a GPU application requires a dynamic array or similar solution, many times it implements an specific and hand-tailored structure that suits the application needs. A common example is when working with triangular meshes \cite{Hatipoglu1997ParallelTM,Mousa2021HighperformanceSO}. In the first work authors introduces a general idea, using parallel prefix-sum to obtain the indexes at which each threads inserts an element. On the other hand, the second work instead of dealing with dynamic memory, introduces handles to each of the graphs elements in a way that modifying the handles offers a similar result to managing dynamic memory.

Given these works, it is clear that a generic vector-like structure is still missing in GPU programming. Specially one with the capacity of being resized dynamically, competitive in performance, and able to take advantage of asynchronous parallelism as well as adapt to the GPU architecture. In this work, we focus on studying the growing aspect of such desired structure.

\section{Current Known GPU Approaches}
\label{sec:current-approaches}
The design idea for a parallel growable array can be divided into two parts. The first part consists of the data structure, as well as how it is resized when needed and the second part consists of how elements are inserted in the array. 

\subsection{Known GPU Data Structure Approaches}
Two approaches are known for growable GPU arrays:

\subsubsection{Static}
The static data structure consists of a flat C like array allocated with \texttt{cudaMalloc} at the start of the program and insertions can occur in a GPU kernel by each thread using a parallel insertion algorithm. This approach does not support any kind of resize operation and it is necessary to know the maximum possible size beforehand for it to not result in a segmentation fault. For many GPU applications, this is the default way of managing a dynamic growth of memory.

\subsubsection{Semi-static}
Similar to the static approach, a C array is utilized as the base structure, but with mechanisms to grow its size via memory reallocation and synchronization from the Host, which allows to use resizing schemes such as the one from doubling arrays. Another possibility is to use a low-level API for virtual memory management provided by CUDA \cite{Perry2020LowLevelMM}, which allows skipping the data copy between doubling arrays. The low-level CUDA API offers functions to modify the mappings between virtual and physical memory. This allows to allocate only the desired extra memory and remap the virtual memory in such a way that indexing is contiguous even if the elements are not physically contiguous. It is worth mentioning however that this benefit comes at the cost of some fragmentation in GPU memory.

\subsection{Known GPU Parallel Insertion Approaches}
The main objective of the parallel insertions algorithms is to allow multiple threads to insert data and consistently update the global size of the array. This involves giving each inserting thread a unique index greater than the previous size and less than the subsequent size, such that each thread inserts its element in a different position as if it was a contiguous array. Three approaches have been identified for GPU architectures.

\subsubsection{Atomic}
The simplest way of obtaining a unique position for each new element is to use the CUDA instruction \texttt{atomicAdd}, which takes as parameters a memory address and an addend. It returns the value stored in the address and updates its value by adding the addend. For the insertion algorithm each inserting thread adds 1 to the size of the array, obtaining an index where to insert the element and updating the size of the array.

\subsubsection{Parallel Prefix-sum}
A more parallel approach for inserting elements consist of considering numbers of insertion per thread as an array with 0s or 1s depending if the threads need to insert an element and calculating the prefix-sum of this array. In CUDA this can be implemented locally per block with the warp \texttt{\_\_shfl\_up\_sync} instruction and globally with atomic operations.

\subsubsection{Tensor-cores Parallel Prefix-sum}
As demonstrated by Dakkak et al. (2019) \cite{Dakkak2019AcceleratingRA} it is possible to further accelerate the prefix-sum computation with CUDA tensor cores by representing the work as matrix multiplications. Although this approach works better on dense problems with a many-to-one mapping between data and threads, respectively, it is still a motivation to consider it in this work and know how efficient is this new use case for tensor cores outside machine learning.

All these known approaches are taken into account when proposing a new data structure for a growing array fully on GPU. In terms of data structure, the goal is to have a fully dynamic one that can update during kernel execution (static and semi-static cannot, therefore their role is for comparison purposes), and for the insertion approach all three are evaluated in order to choose the fastest one to be used in the new GGArray structure.

\section{Proposing GGArray}
\label{sec:ggarray}
In this section we introduce GGArray, a fully dynamic GPU structure with an interface similar to an array based on the LFVector dynamic data structure which was originally proposed for CPU architectures. The LFVector is based on the idea of doubling arrays, where the size of the array is doubled whenever more space than its current capacity is needed. This doubling is usually done by creating a new array with double the size and moving all elements of the previous array before deleting it. However, when multiple threads are accessing the elements of the array at the same time, having two copies introduces a synchronization problem as it is necessary for all threads to know when the array changes. The LFVector eliminates this problem by dividing the array into non-contiguous blocks, each double the size of the previous one, and allocating them when needed.

\begin{figure*}[ht!]
\centering
\includegraphics[scale=0.2]{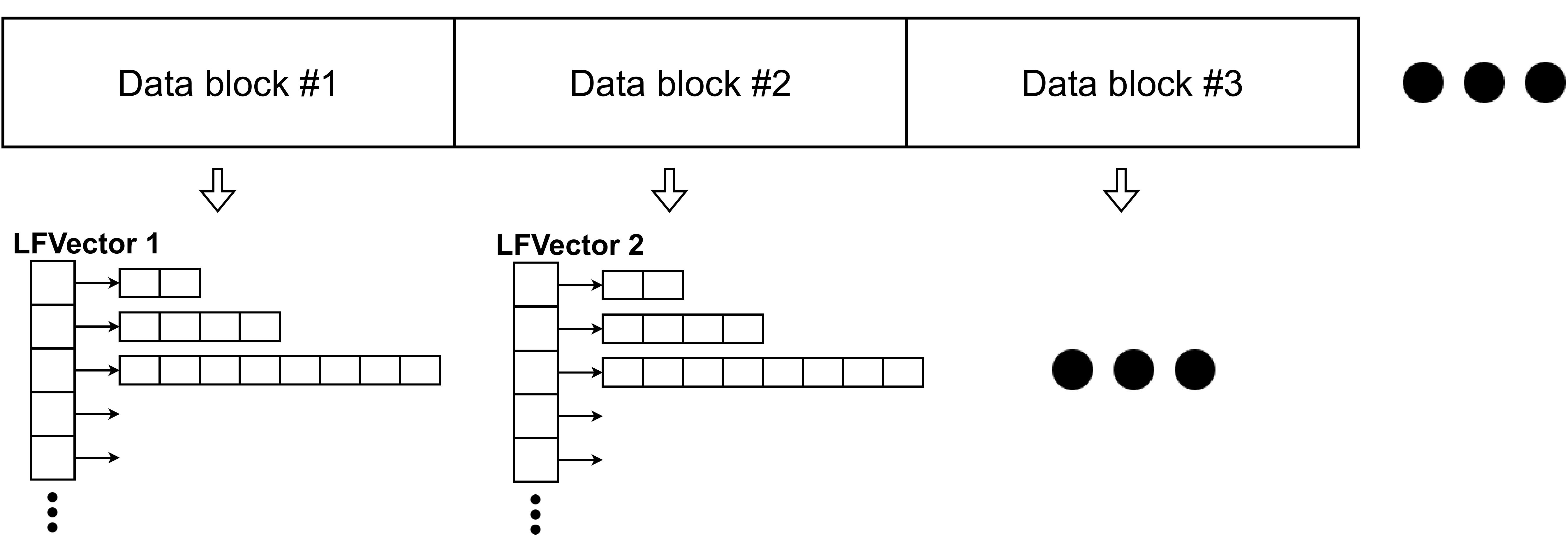}
\caption{In GGArray, one LFVector per data block is used. This allows to correlate each data block with a thread block independently.} \label{figLFVector}
\end{figure*}

The original idea of the LFVector lies on the use of Compare-And-Swap (CAS) with every thread trying to allocate memory and deleting all except from the first allocated memory. This mechanism is not suitable for GPUs, given that there is not enough memory for thousands of threads to try allocating a doubled buffer before knowing which one succeeded. Another way would have been to use the busy waiting logic for the threads to synchronize when allocating memory, but this approach only works inside a block, where there is a high synchronization between threads, and not among blocks. For multiple thread blocks deadlocks can occur as the swapping between busy and idle blocks can make the allocating block become swapped out by the busy ones, locking the execution for an unknown amount of time. There are two ways to solve this problem, the first is globally synchronizing all the blocks and the other is avoiding the synchronization by dividing the data structure. We opted to further divide the array and take advantage of the block independent execution by creating a macro structure made of multiple LFVectors, one for each block of threads as illustrated in Figure \ref{figLFVector}. 

\begin{figure}[ht!]
\centering
\includegraphics[scale=0.7]{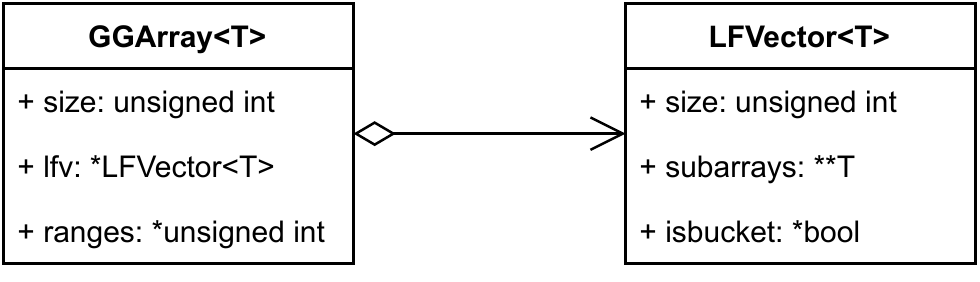}
\caption{GGArray structure, each LFVector maps to a GPU block and it is independent from other LFVectors.} \label{figClassDiagram}
\end{figure}

Although this design limits the parallelization of the problem as there is a fixed number of threads that can work on each block, it does not add additional limits to the number of blocks neither the amount of data to be processed in each block (i.e., thread coarsening can be applied so that more work is assigned to each thread). Also the amount of GPU cores is increasing in recent GPU architectures (Ampere, Lovelace), reaching the tens of thousands, therefore the larger the array of LFVectors, the less noticeable this parallel limit will be perceived. This design also allows to synchronize LFVectors with builtin CUDA instructions. The formal diagram for the structures is shown in Figure \ref{figClassDiagram}. 

The functionalities are shown in Algorithms \ref{alg:pushback} and \ref{alg:newbucket}.

\begin{algorithm}
\caption{LFVector push\_back}
\label{alg:pushback}
\begin{algorithmic}
	\Require e
   \State $idx = get\_insertion\_index()$
   \State $b = get\_bucket(idx)$
   \If {$bucket[b] = nil$}
   \State $new\_bucket(b)$
   \EndIf
   \State $synchronize()$
   \State $vector(idx) = e$
\end{algorithmic}
\end{algorithm}

\begin{algorithm}
\caption{new\_bucket for an LFVector}
\label{alg:newbucket}
\begin{algorithmic}
	\Require b
   \If {$not\  CAS(isbucket(b), False, True)$}
   \State $bsize =  2^{log(first\_block\_size) + b}$
   \State $bucket[b] = malloc(bsize * type\_size)$
   \EndIf
\end{algorithmic}
\end{algorithm}

Given that each LFVector is constrained to its own block, the dynamic array requires a new structure to keep track of its size and the ranges encompassed by each LFVector. This structure is a prefix-sum of the sizes of all LFVectors, it contains the index of the first element contained by them. We are using a C-style array, which offer great amount of parallelism for updating its values, but it is needed to search over this array to locate the LFVector that contains a certain index. Using a prefix-sum allows us to partially reduce the time needed for this search using binary search.

The insertion method for the GGArray consists of  delegating the process to each of the LFVectors. At a global level, it only needs to update the global size and prefix-sum indices, as the actual \texttt{push\_back} insertions are taken care locally by the LFVectors during kernel execution.

\section{Theoretical Memory Usage}
\label{sec:theo-mem-usage}
\begin{figure}[ht!]
\centering
\includegraphics[scale=0.5]{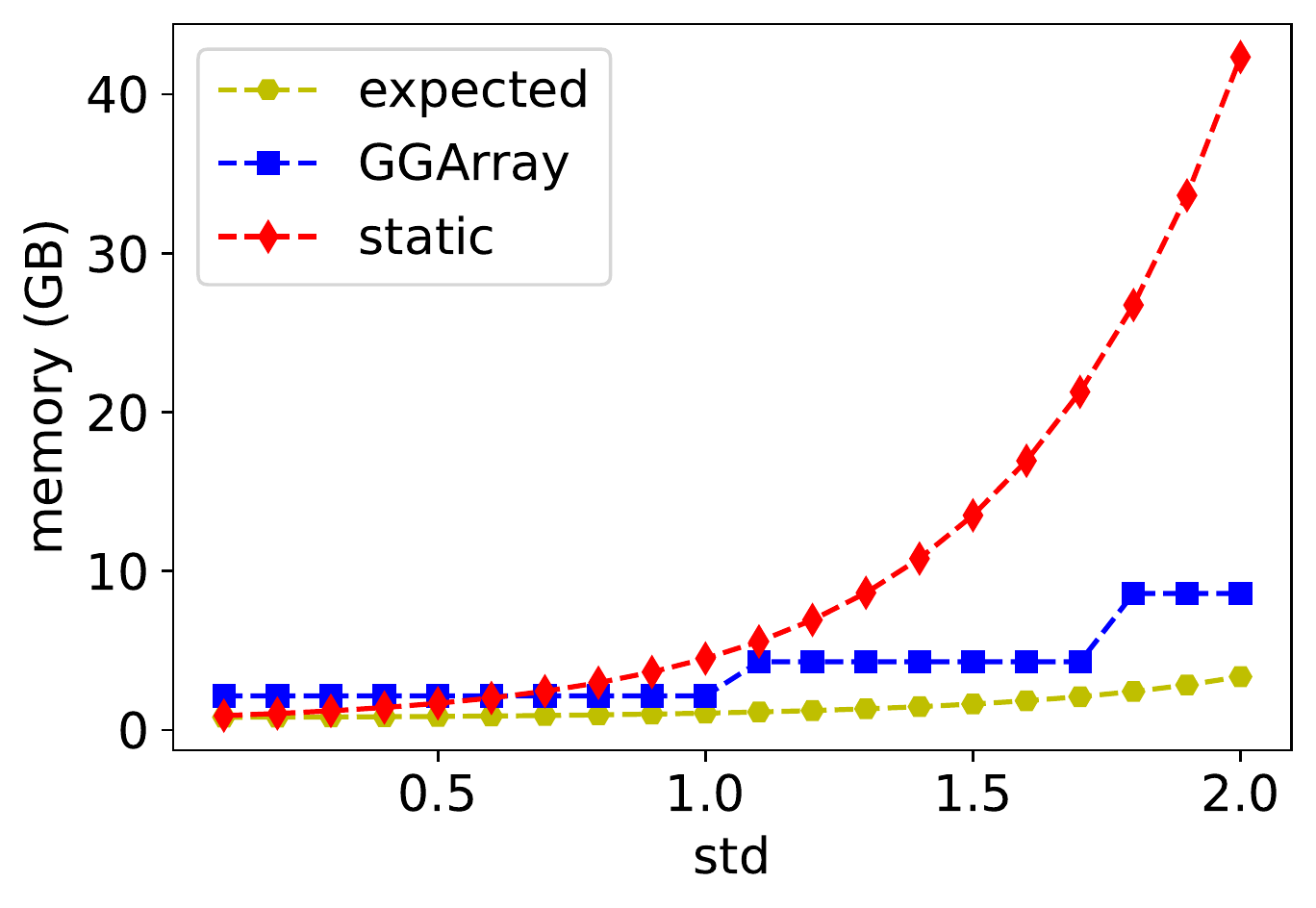}
\caption{Theoretic memory usage of GGArray and the static/semi-static arrays.} \label{figTheoreticMem}
\end{figure}

A major advantage of the GGArray is its ability to dynamically grow during kernel execution according to the needs of the application. This allows programmers to run applications without concern about the amount of memory to pre-allocate nor if the program will fail due to an invalid memory address. This is not a big issue for static methods when it is known beforehand the insertion behaviour of each thread, however, when there is not enough information or there is only a rough idea of the growing behaviour of the array, the worst case for the static or semi-static methods start to grow excessively. Figure \ref{figTheoreticMem} shows the memory needed for an example where the amount of insertions are given by the size of the array times a factor given by a log-normal distribution with parameters $ \mu = 0   $ and $ \sigma \in [0, 2] $. 

It shows how with a larger standard deviation and uncertainty about the amount of insertions realized, more memory it is needed for the static method to fail only 1\% of the times it is executed. On the other hand, GGArray remains closer to the optimal amount of memory needed, reaching in the worst case approximately $2\times$ more.

\section{Experimental Evaluation}
\label{sec:exp-eval}
All performance tests were ran on the TITAN RTX and A100 GPUs. Their specifications are listed in Table \ref{table:specs}.

\begin{table}[ht!]
\caption{GPUs specifications}\label{table:specs}
\centering
\resizebox{\columnwidth}{!}{
\begin{tabular}{|l|l|l|} 
\hline
                 & TITAN RTX    & A100          \\ 
\hline
CUDA Cores       & 4608         & 6912          \\ 
\hline
Tensor cores     & 576          & 432           \\ 
\hline
Memory           & 24 GB        & 40 GB         \\ 
\hline
FP16 performance & 32.62 TFLOPS & 77.97 TFLOPS  \\ 
\hline
FP32 performance & 16.31 TFLOPS & 19.49 TFLOPS  \\ 
\hline
Base Clock Speed       & 1350 MHz     & 765 MHz       \\
\hline
\end{tabular}
}
\end{table}

\subsection{Choosing the Fastest Insertion Algorithm}
\begin{figure*}
\centering
\includegraphics[scale=0.4]{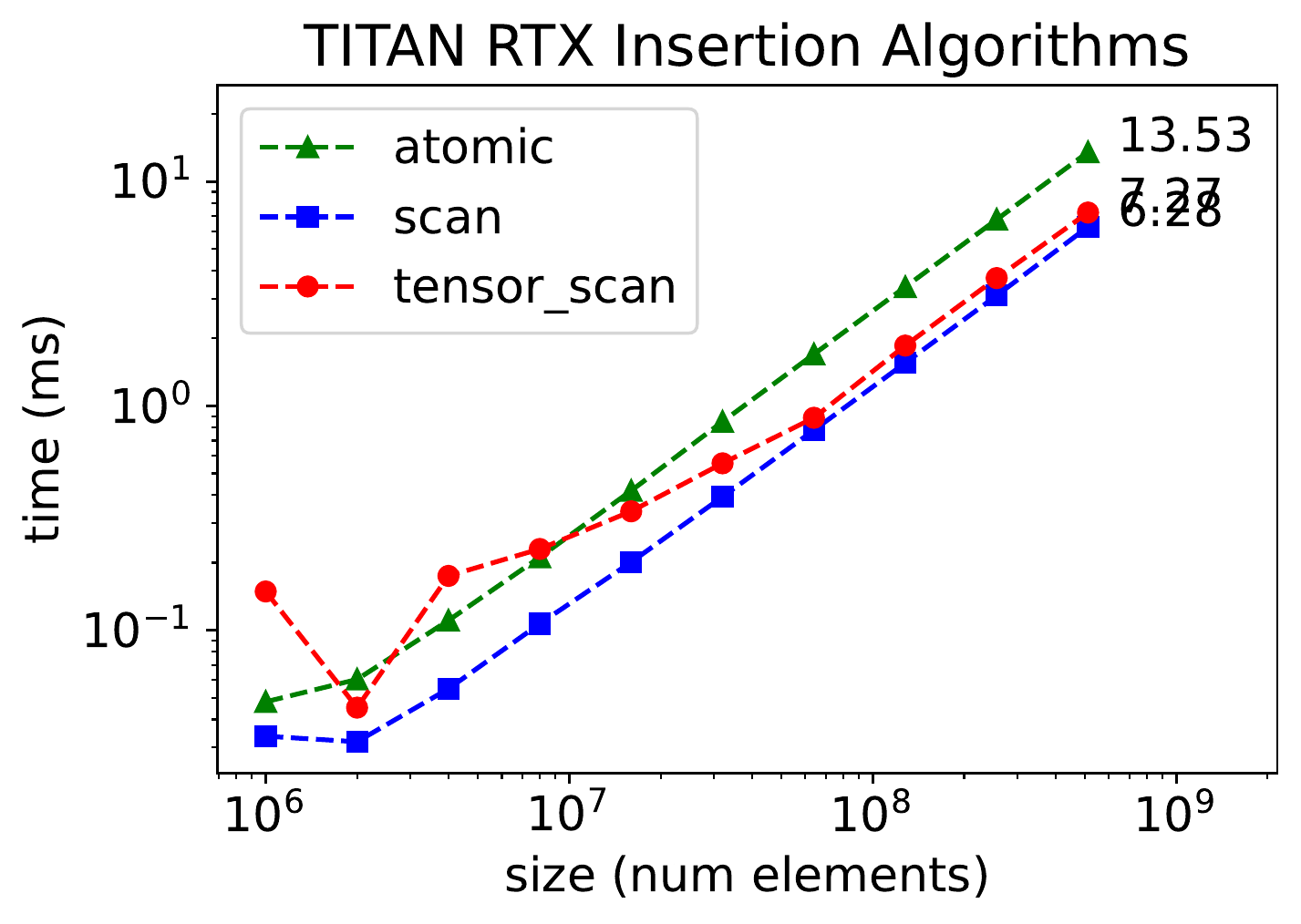}
\includegraphics[scale=0.4]{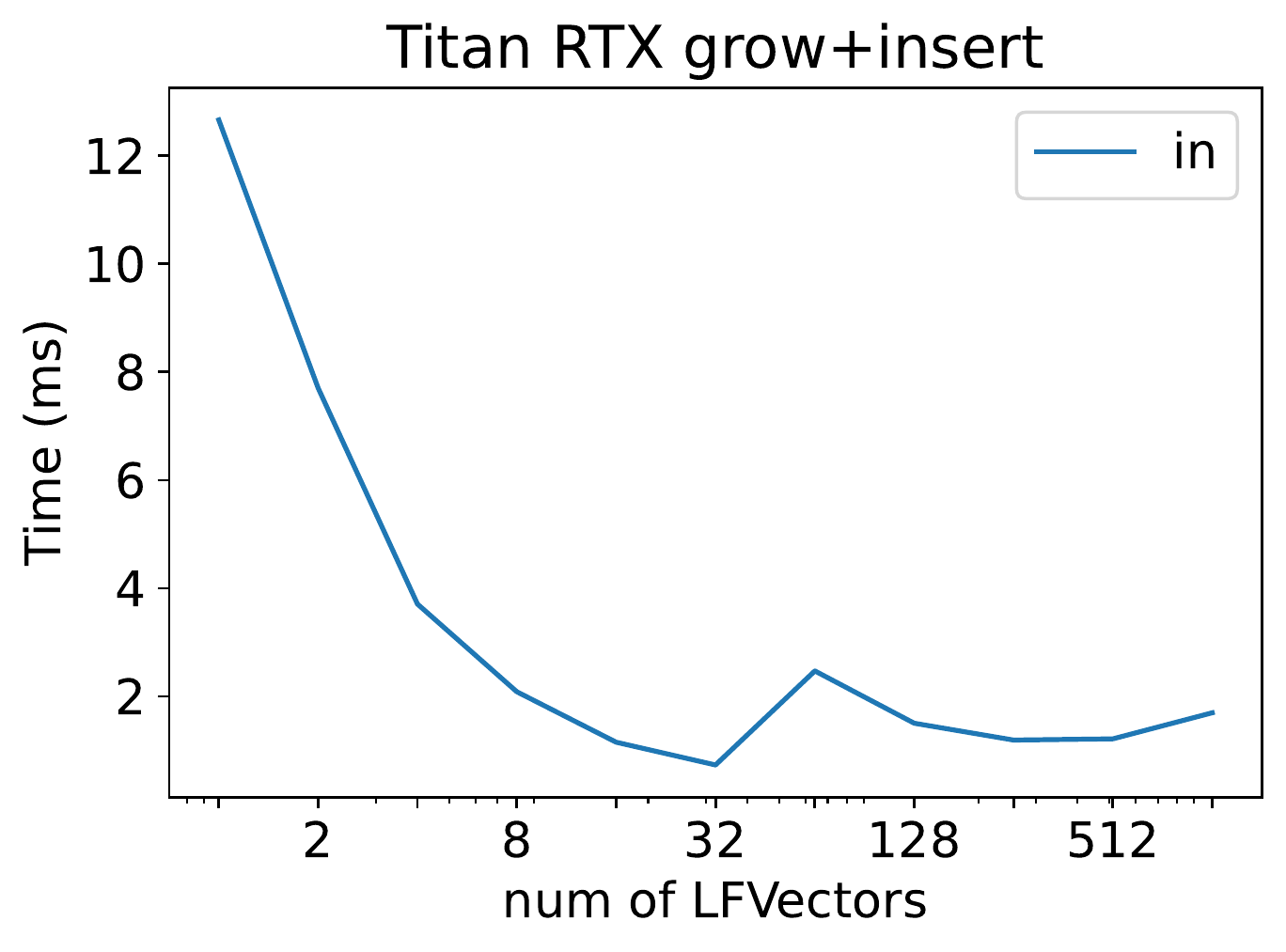}
\includegraphics[scale=0.4]{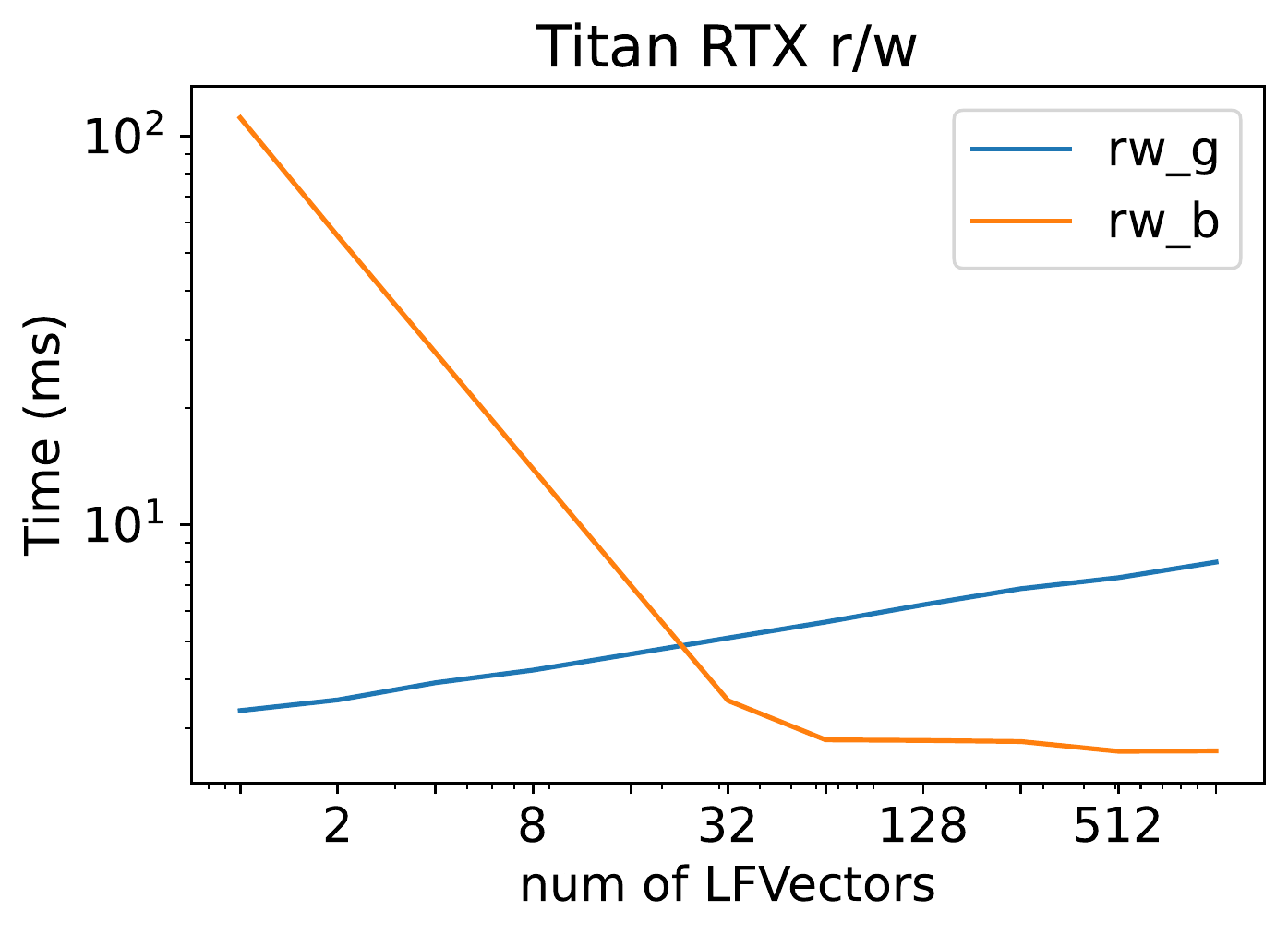}\\
\includegraphics[scale=0.4]{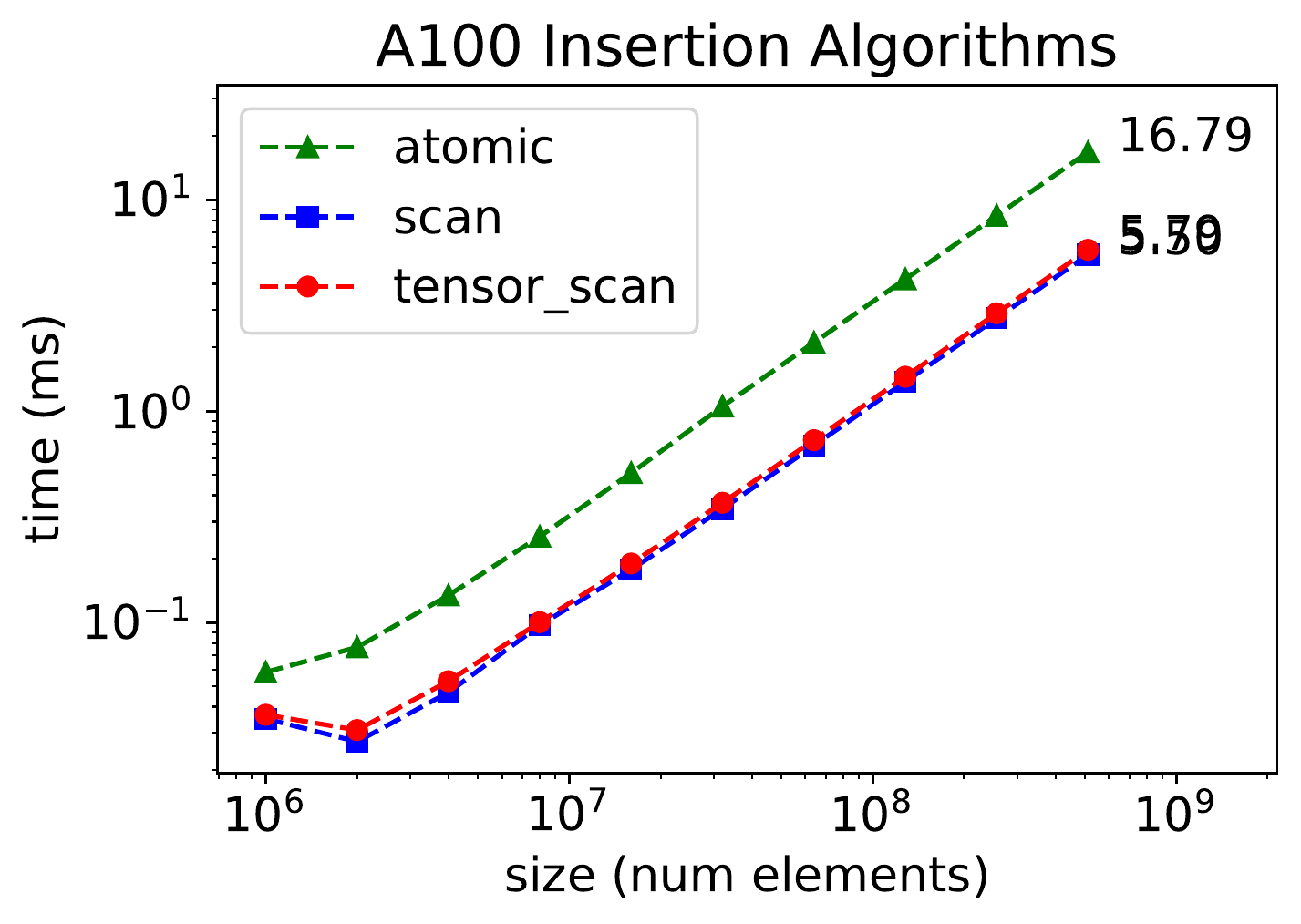}
\includegraphics[scale=0.4]{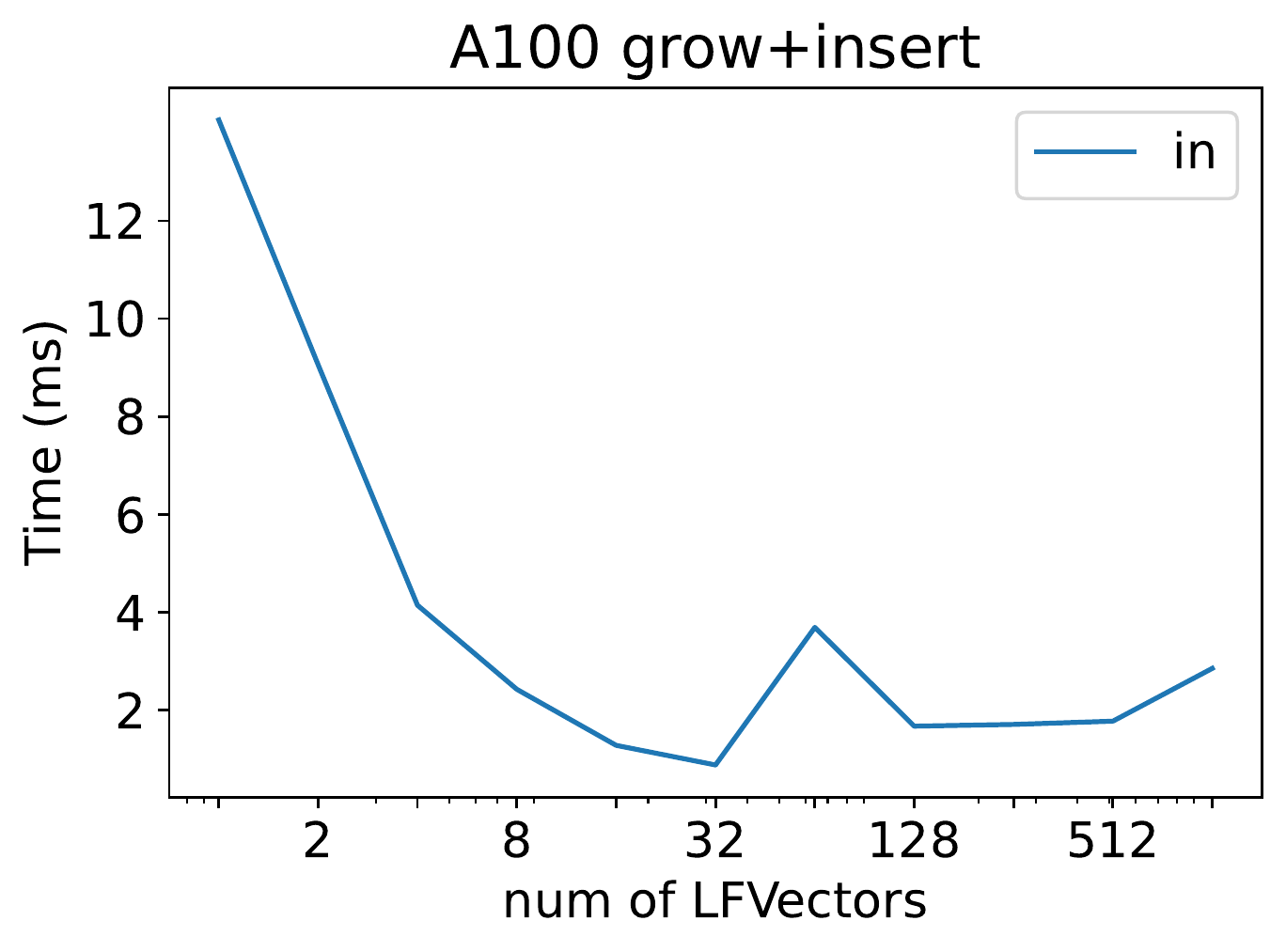}
\includegraphics[scale=0.4]{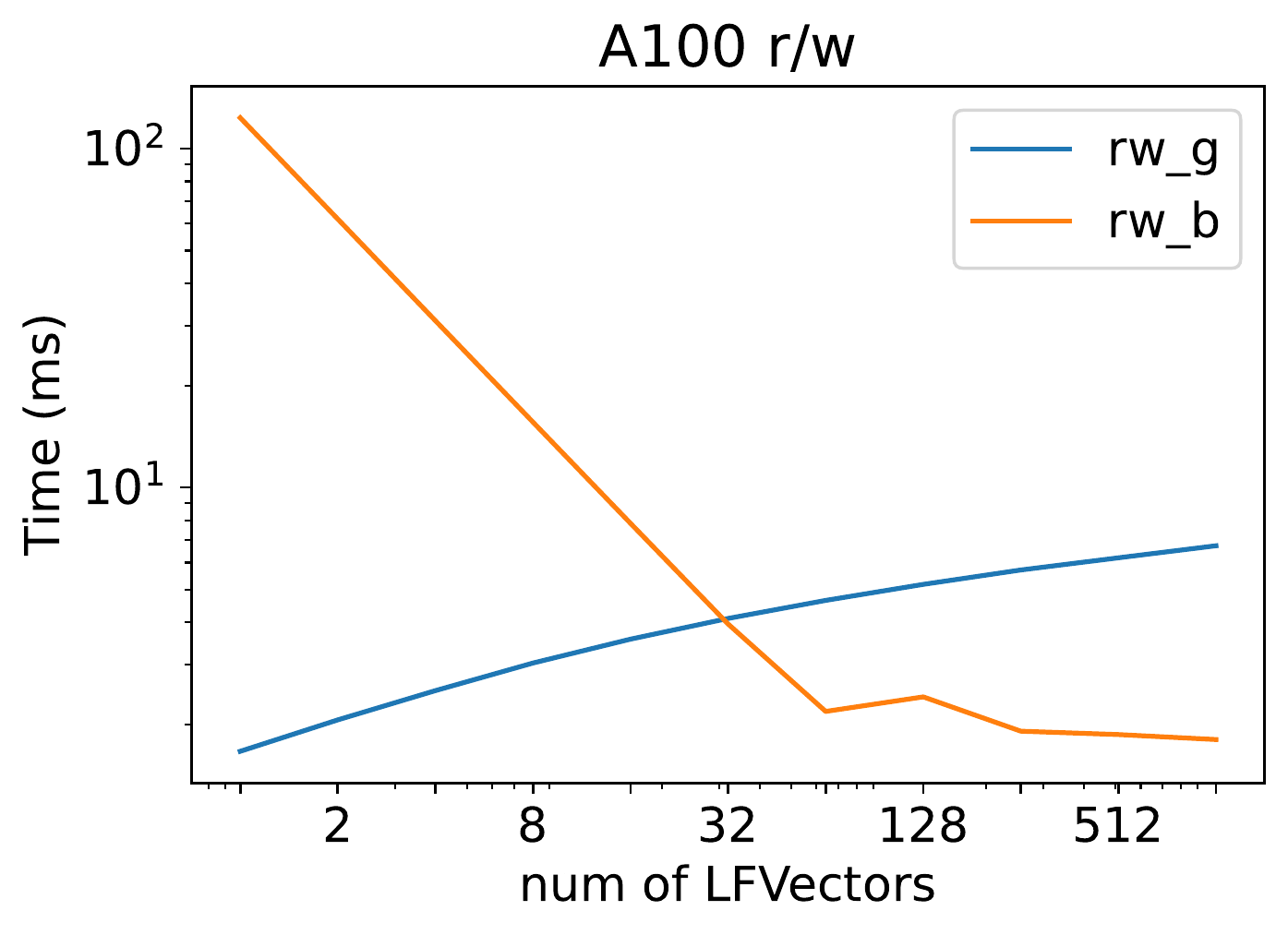}
\caption{Insertion, grow$+$insert and r/w times over size and number of LFVectors.} 
\label{fig:algs_best_block}
\end{figure*}

The static array was used as the structure to test the different insertion approaches. The reason for not testing with the other structures is that insertion algorithms are independent of the underlying structure and the static array is the simplest, allowing to only measure the time of the insertion algorithm without being affected by the time needed to access the structure elements. The test consists of an array with $1\mathrm{e}{6}$ elements and a sufficient capacity for duplicating its size 10 times, finishing with an array of $1.024\mathrm{e}{9}$ elements. Time is measured for each iteration of duplication. Figure \ref{fig:algs_best_block} (first column) shows the results for the algorithms using atomic operations, warp-shuffle prefix-sum and tensor-core prefix-sum. Insertions with atomic operations were the slowest, while the shuffle scan is the fastest closely followed by the tensor core one.

Regarding the scan operation being slower when implemented with tensor cores than with the usual algorithms, as opposite from other studies in the state of the art, it is due to not meeting the necessary workload for this specific case. For this particular test, the size of the problem for the insertion algorithm is the amount of threads participating in the insertion. Thus, when using tensor cores that multiplies 16x16 matrices per warp, there are not enough elements to fill all matrices from all warps. In the tensor scan algorithm only one eighth of the warps are realizing the algorithm while the rest are idle, not taking advantage of the full potential of tensor cores. Other applications with a higher ratio of data elements to threads could exhibit an scenario where the tensor core approach runs faster. It is also important to note that the difference between the two scan versions is lower in the A100 GPU, due to the improvement in tensor cores from the previous generations being larger than the improvement in CUDA cores.

\subsection{Choosing An Optimal Number of LFVectors}
The variables that affect time execution of the GGArray are its size, the amount of blocks in which it is divided and the amount of memory allocations previously realized. The size impacts read/write and insertion operations since the more elements the array contains, more operations are needed to operate over the whole array. Similarly, more concurrent blocks allows a larger amount of parallelization in these operations, except for atomic ones. In the case of memory allocation more parallelization means more allocations which do not occur in parallel due to the limitations of current technology.
We ran tests to determine the optimal amount of blocks, with the results shown in Figure \ref{fig:algs_best_block} (second and third columns). The Figure shows the amount of time it takes to duplicate the amount of elements in the array utilizing different numbers of blocks. The duplication process includes the memory allocation and insertion of elements. The plots of the third column show the time spent to realize read/write operations in two ways. The first one (rw\_g) utilizes the structure as if it were an array with one thread per element. On the other hand, rw\_b follows the block structure and uses one GPU block per array block avoiding the process of determining which block contains an element which is slow. In general, a low number of blocks implies the growth of the structure is slower due to the lack of parallelization in insertion, reaching optimal configurations at 32 and 512 blocks. With over 32 blocks, read/write operations by block are faster and their time is inversely related to the number of blocks.

\subsection{Growable Array Operations}
\begin{figure*}
\centering
\includegraphics[scale=0.4]{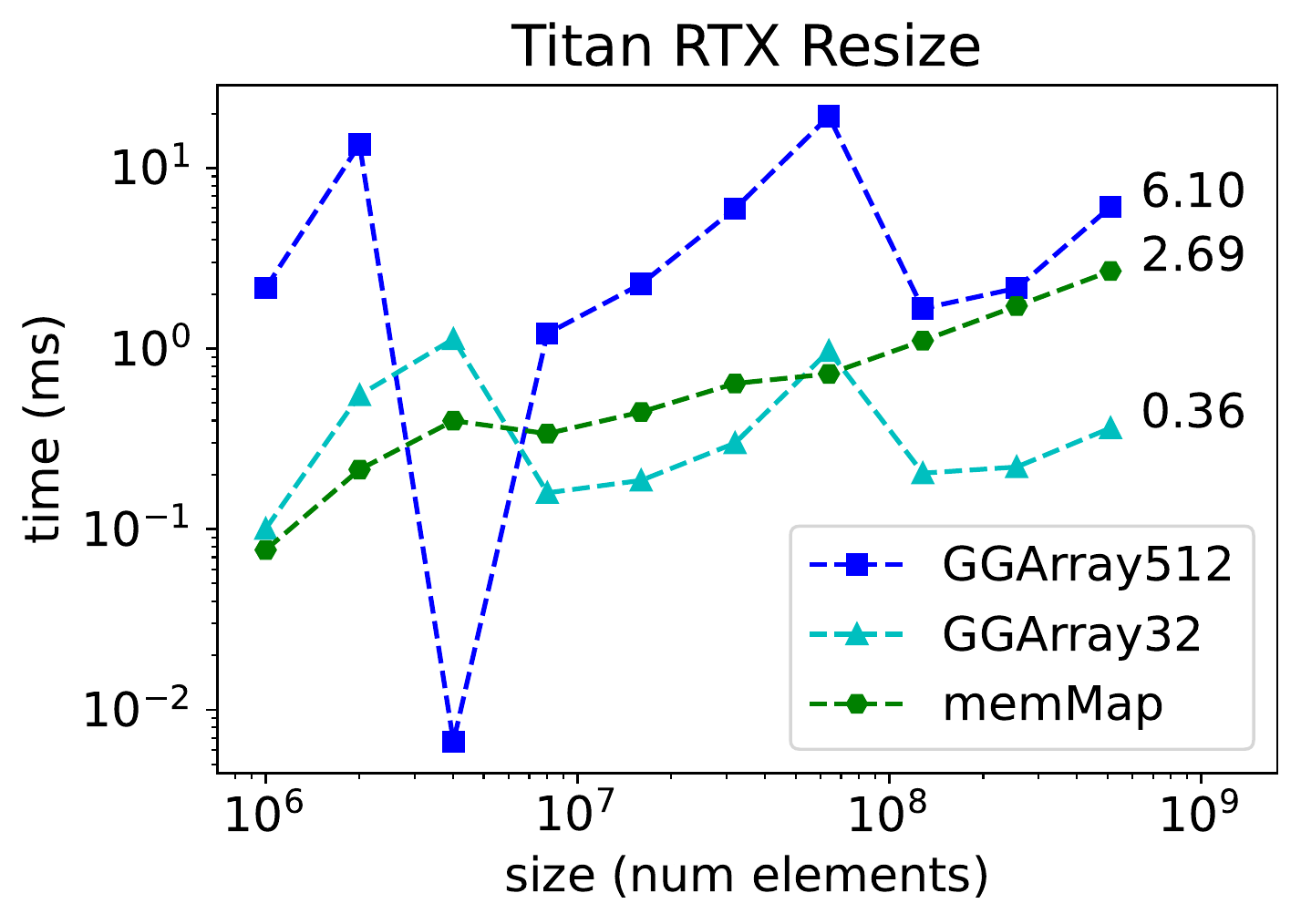}
\includegraphics[scale=0.4]{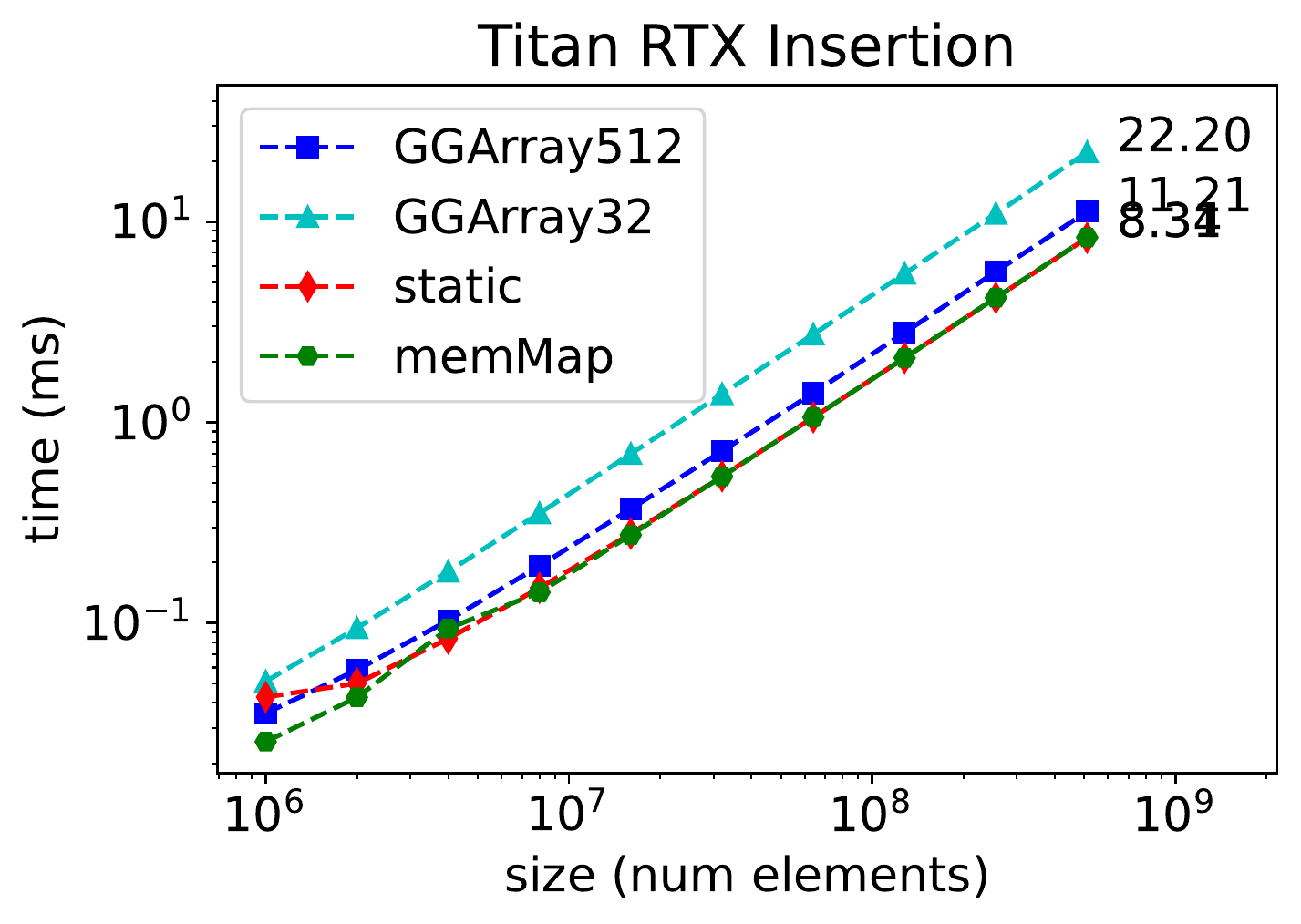}
\includegraphics[scale=0.4]{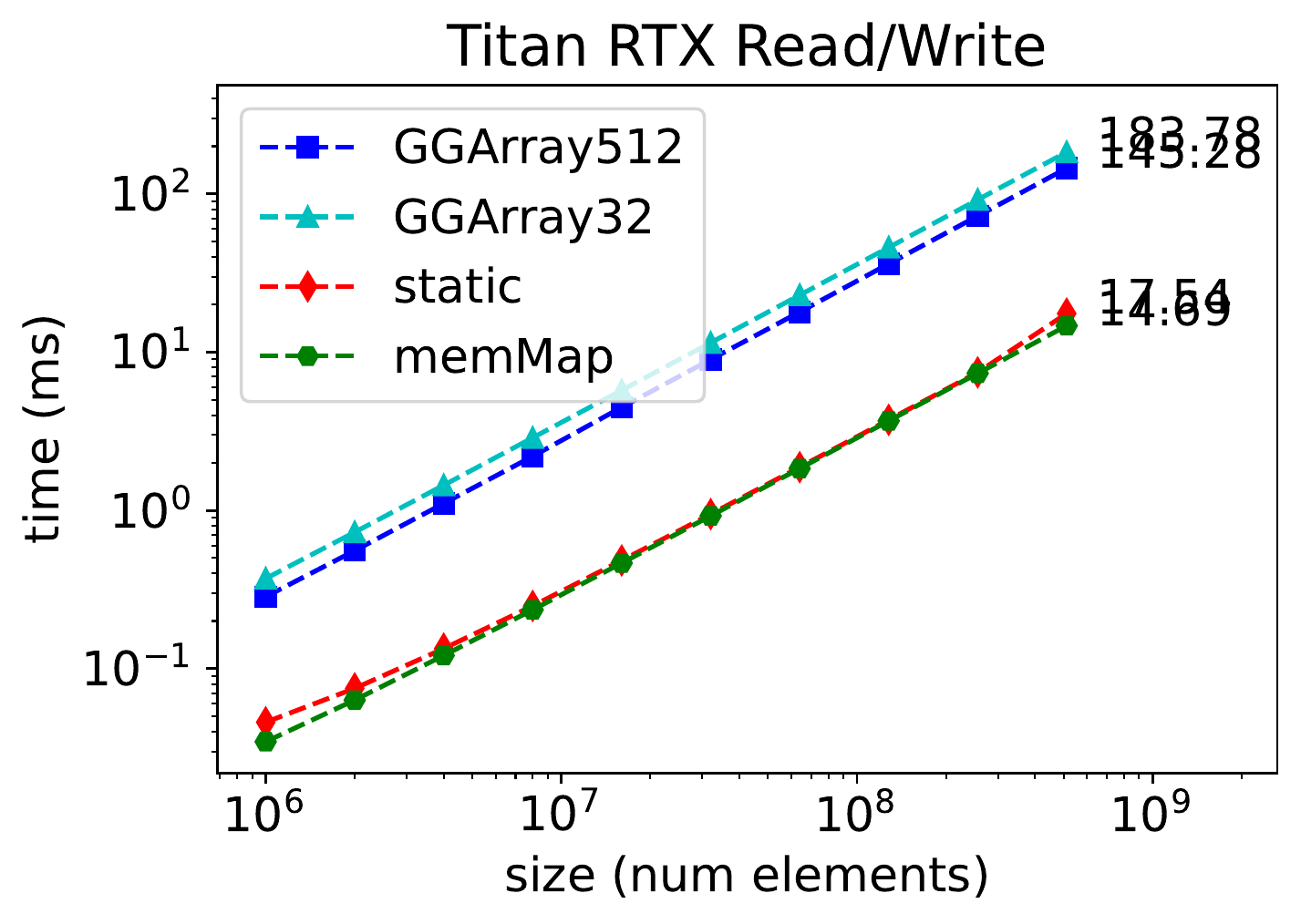}\\
\includegraphics[scale=0.4]{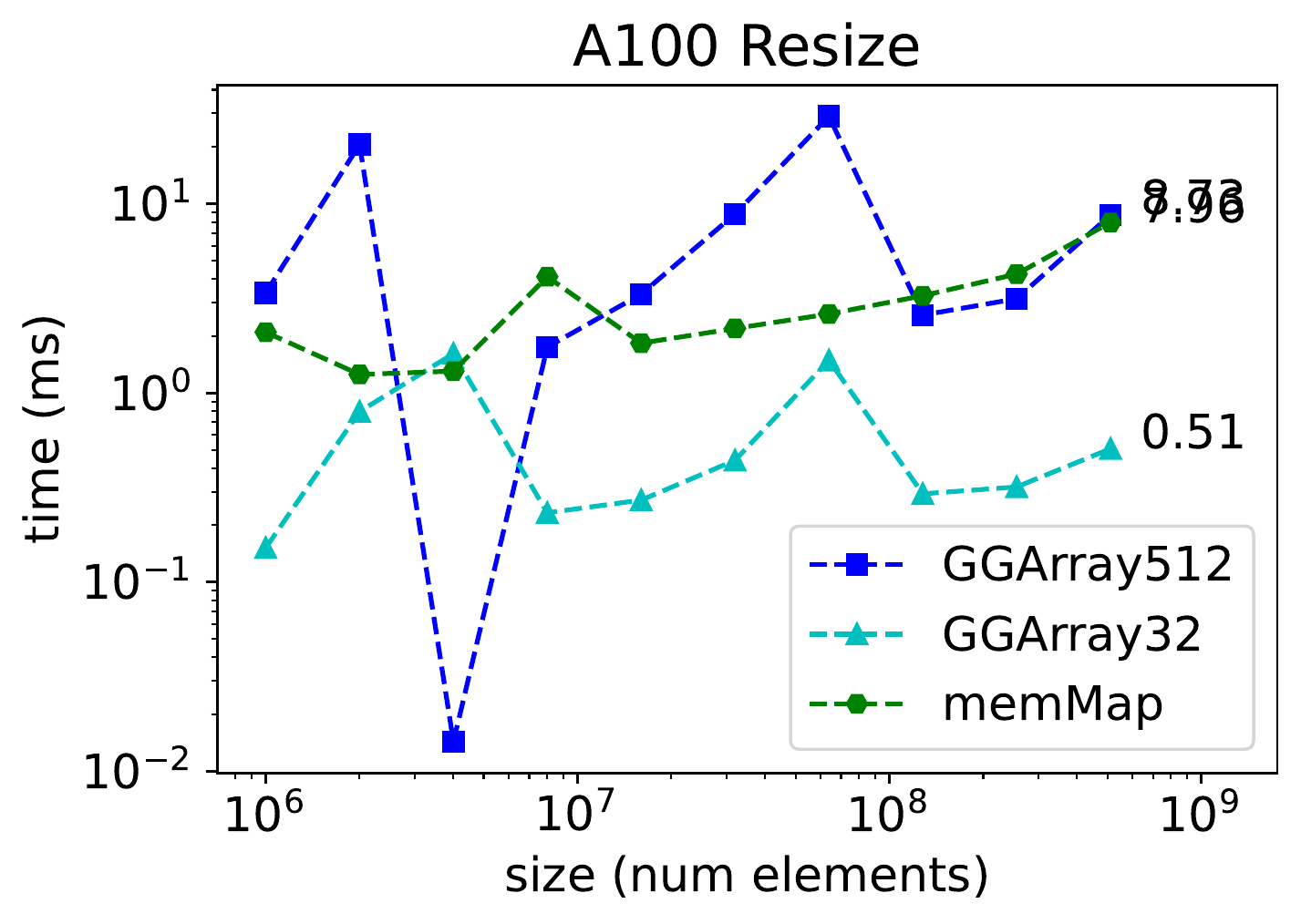}
\includegraphics[scale=0.4]{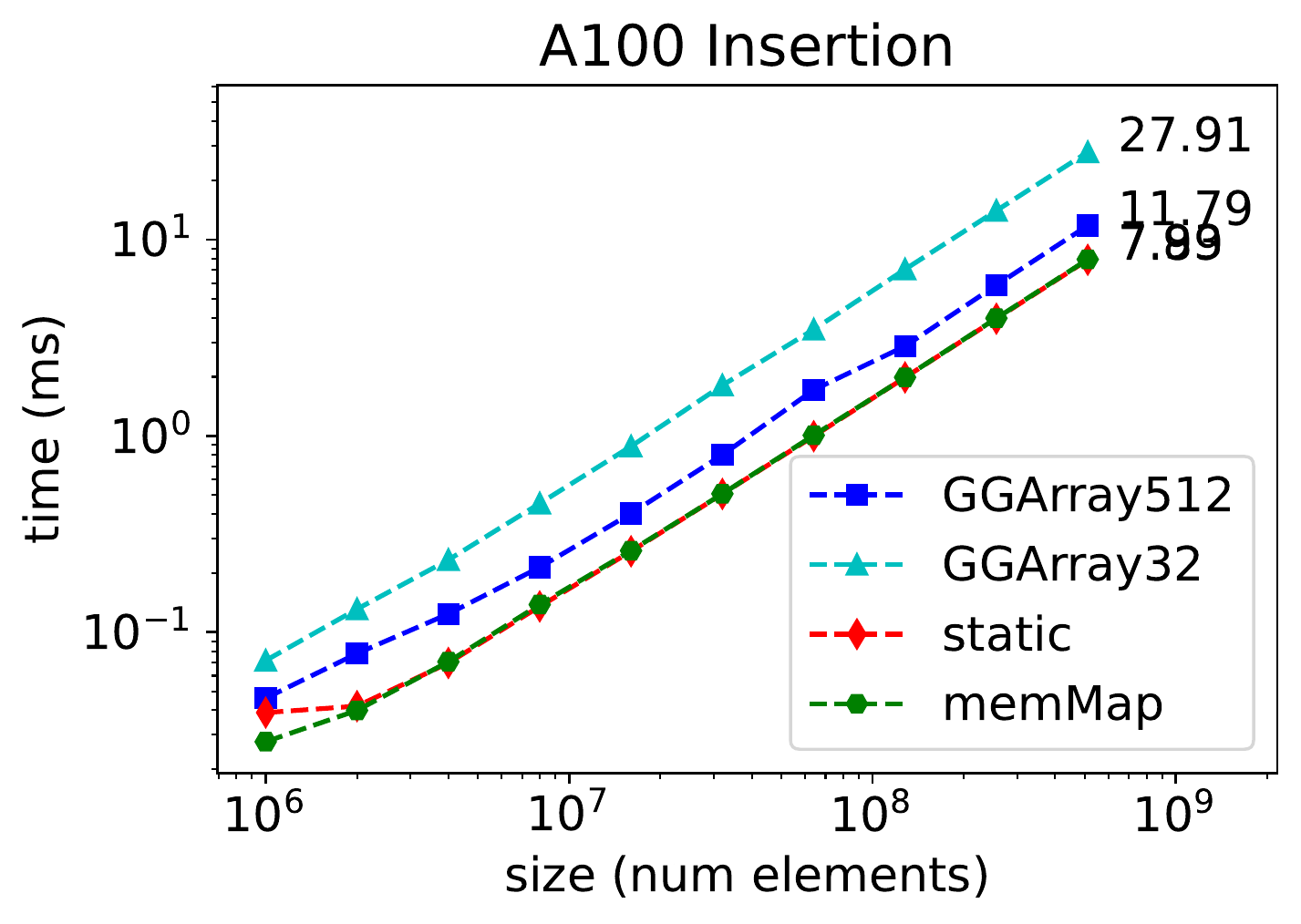}
\includegraphics[scale=0.4]{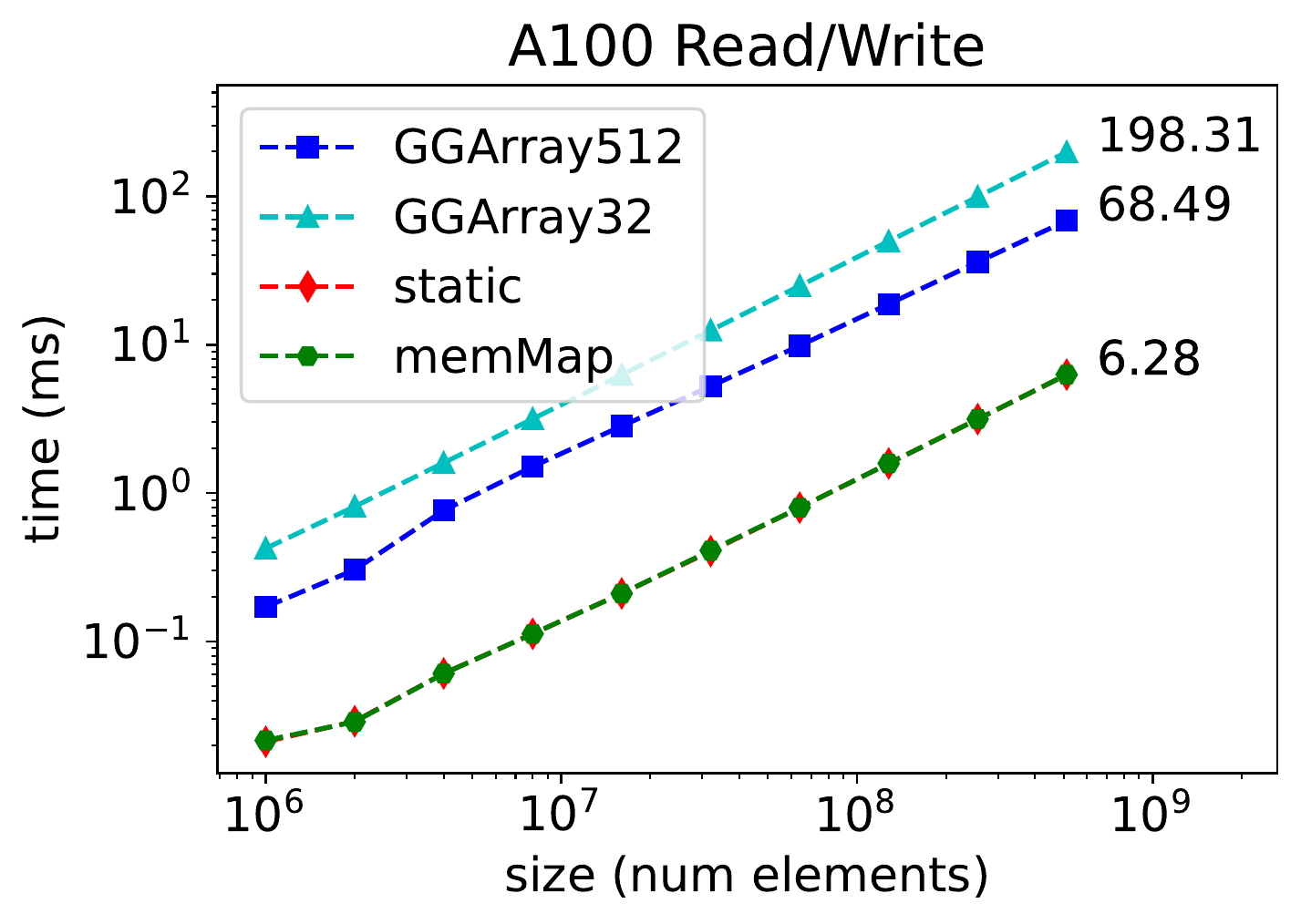}
\caption{Time of operations to duplicate array size each iteration stating with a size of $1e6$. \emph{Resize} increases the capacity if necessary, \emph{insertion} inserts one element per each previous element and \emph{read/write} performs an operation per each element in the updated array} 
\label{fig:results}
\end{figure*}

The experiment to test the performance of array operations consists of starting with an array of size $1\mathrm{e}{6}$ and duplicating (with scan algorithm) its size 10 times. Inserting less elements than the size of the array doesn't reduce the time taken, because even threads that do not insert elements play a role in the insertion algorithm and are also needed for synchronization. The duplication of the array is divided in the grow operation and the insertion operation. Also, for each size the time to operate on each of its elements is measured. The results are displayed in Figure \ref{fig:results}. Moreover, Table \ref{table:results_last} shows the exact time taken by each operation on the last iteration. In accordance to the previous results, 32 and 512 blocks are utilized for read/write operations per block. In the legend, GGArray32 and GGArray512 correspond to the variants of the proposed structure with the numbers of blocks in which it is divided, and \texttt{memMap} is the semi-static array using the NVIDIA low-level memory management API. The first two figures show the time to duplicate the capacity of the arrays. The two in the middle depict the time needed for the insertion of elements filling the capacity of the array. The last column of plots displays the time required to realize operations in all elements of the array. The operation used is a kernel that adds $+1$, $30$ times to each element.

\begin{table}[ht!]
\centering
\caption{Time (ms) to duplicate an array of size $5.12\mathrm{e}{8}$ in the last iteration using NVIDIA A100}\label{table:results_last}
\resizebox{\columnwidth}{!}{
\begin{tabular}{|l|l|l|l|} 
\hline
        & grow  & insert & read/write  \\ 
\hline
static  &  ~ ---     & 7.07   & 6.27        \\ 
\hline 
memMap  & 5.21  & 7.87   & 6.28        \\ 
\hline
GGArray512 & 8.76  & 11.79  & 69.73       \\ 
\hline
GGArray32  & 0.52 & 27.90  & 198.32      \\
\hline
\end{tabular}
}
\end{table}

It draws attention that the third resize barely takes time. The explanation is that the growth in capacity of the GGArray is not a constant factor, but it tends to two as the size increases, in this case no resizing took place because the capacity from the previous iteration was enough. The major drawback of the proposed structure are the slow read/write operations. While allocating memory for a large amount of LFVectors and inserting elements are slower than the other structures the difference is not large enough to cause a bottleneck, especially when realizing more complex operations in-between resizing. However, in order to realize application work with the contents of the structure it is necessary to read and write its elements, and these operations are slow, currently more than 10 times slower, even when working by block without the need to search which LFVector contains each element. This is produced by the more complex indexing operation, a worse cache locality and the need to pass over multiple pointers to reach an element. Something that may only be resolved by a truly contiguous array.

Still, there are some applications that may benefit from the GGArray structure. For applications that need a dynamic array and that do not have a way to confidently know beforehand the maximum size, or the uncertainty of the maximum size is big enough, our structure offers a way to dynamically grow the array from inside the kernel and using no more than double the necessary memory. Also, applications that can be defined in phases where one phase only inserts elements and the other phases realize work on a static structure could be benefited by moving the elements between our structure and a static array. This reduces the read/write operations to only a few per each growth phase and can still take advantage of the characteristics of static arrays in work phases. Applications that meet these conditions may be encountered in computer geometry and triangular mesh refinement.

\subsection{Case study: Two Phase Application}

\begin{figure}[ht!]
\centering
\includegraphics[scale=0.55]{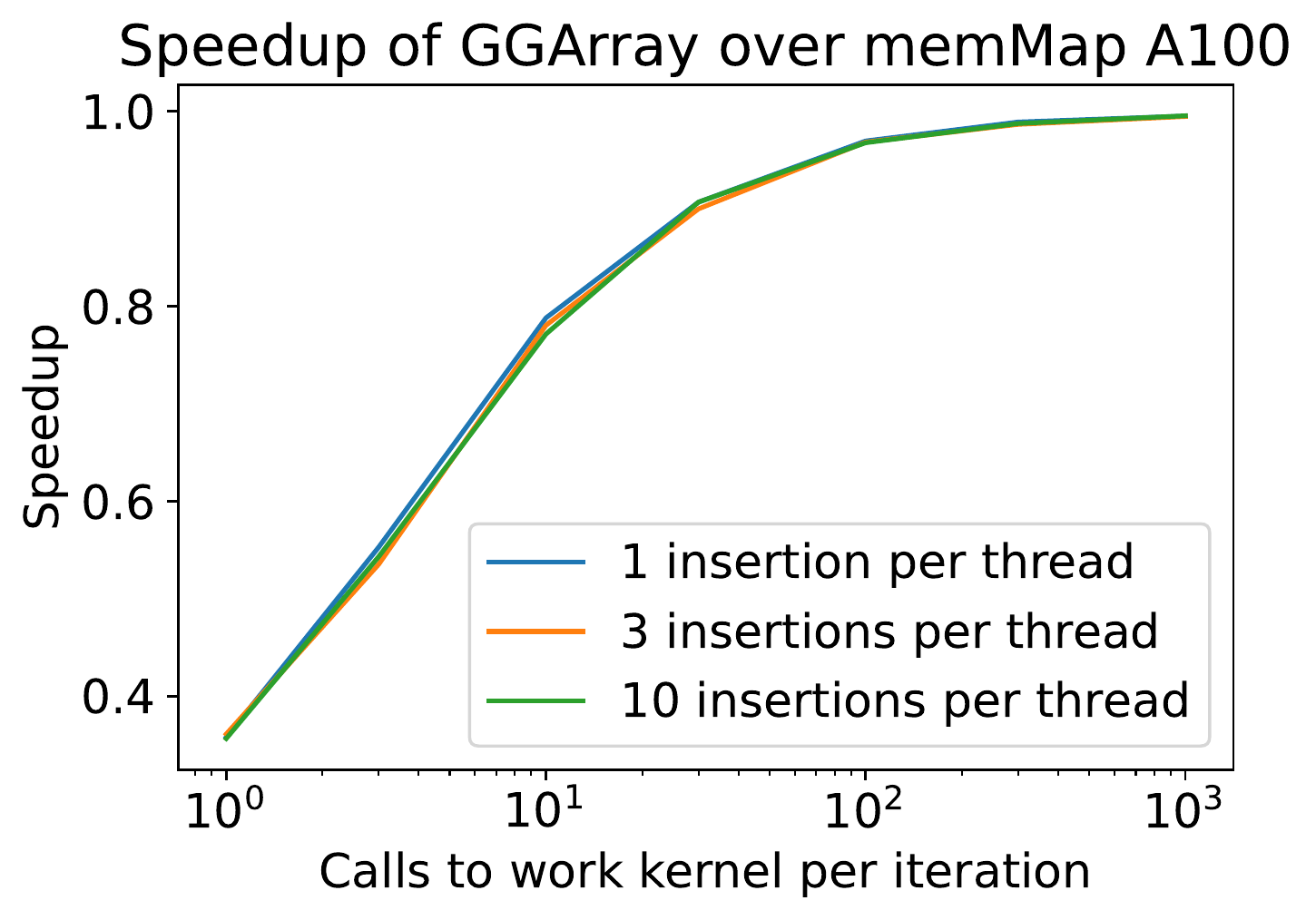}
\caption{Speedup in example use case in application divided in phases} \label{figPhases}
\end{figure}

Figure \ref{figPhases} shows the speedup of GGArray over \texttt{memMap} for a two phase application.
The work phase simply consists of a kernel that adds 1 to every element of the array called multiple times (between 1 and 1000) corresponding to the X axis of the plot. The results illustrates how the overhead added by the dynamic structure can be disregarded when the amount of work on the other phases is big enough. The experiment was designed with 5 repetitions and a starting array size such that after all iterations and independent of the amount of insertions per thread per iteration the final size is $1e9$. Inserting 1, 3, or 10 times the size of the array each iteration does not have an impact on the speedup.

\section{Conclusions and Future Work}
\label{sec:conclusions}
In this work we proposed GGArray, a fully dynamic array for the GPU that offers the interface of a growable array and works inside the GPU without the need of synchronizing through the host. This allows to allocate memory when required from kernel code without the need to pre-allocate all necessary memory. The GGArray has one important drawback; its slow access to the elements, which makes it still unsuitable as a general purpose array. Nonetheless, there are certain applications where it can still prove to be useful, such as where dynamic allocation is crucial, or applications that can be divided into phases where most work is static and the insertion can be done in a separate stage. Also, it is important to note that this structure affects the programming of CUDA kernels to some degree, due to the per-block design which requires threads to stay in execution for warp synchronization and insertion algorithms.

Further improvements are needed for accessing elements faster, one idea that should improve performance significantly is to make GGArray use the L1 cache (programable shared memory) for segments of the array. Also, NVIDIA has made progress in favor of dynamic applications in the lasts years, for example RT cores, and currently they are being researched as a computation tool outside ray tracing. This may be an interesting approach to implement a dynamic data structure. On the other hand the issue of accessing elements doe not exist if a contiguous array is utilized, although it brings a lot of synchronization issues, that could be solved with cooperative groups in combination with the recent thread block clusters introduced with the Hopper GPU architecture. NVIDIA has also recently unlocked the GPU System Processor, a chip similar to a CPU, but inside the GPU. This processor could bring a lot of benefits if it is used for synchronization instead of the CPU. Finally, separating the data structure and allocation from the insertion algorithm leaves open the possibilities for the use of any scan algorithm already studied or even other algorithms that outputs an unique index per thread.

\section*{Acknowledgment}
This research was supported by the Temporal research group, the ANID Fondecyt grant \#1221357 and the Patagón supercomputer of Universidad Austral de Chile (FONDEQUIP EQM180042).

\bibliographystyle{plain}
\bibliography{main}

\begin{thebibliography}{10}

\bibitem{Awad2020DynamicGO}
Muhammad~A. Awad, Saman Ashkiani, Serban~D. Porumbescu, and John~Douglas Owens.
\newblock Dynamic graphs on the gpu.
\newblock {\em 2020 IEEE International Parallel and Distributed Processing
  Symposium (IPDPS)}, pages 739--748, 2020.

\bibitem{BELL2012359}
Nathan Bell and Jared Hoberock.
\newblock Chapter 26 - thrust: A productivity-oriented library for cuda.
\newblock In Wen mei W.~Hwu, editor, {\em GPU Computing Gems Jade Edition},
  Applications of GPU Computing Series, pages 359--371. Morgan Kaufmann,
  Boston, 2012.

\bibitem{Busato2018HornetAE}
Federico Busato, Oded Green, Nicola Bombieri, and David~A. Bader.
\newblock Hornet: An efficient data structure for dynamic sparse graphs and
  matrices on {GPUs}.
\newblock {\em 2018 IEEE High Performance extreme Computing Conference (HPEC)},
  pages 1--7, 2018.

\bibitem{cudaBestPractices}
NVIDIA CORPORATION.
\newblock {CUDA} {C++} best practices guide 11.6.1, 2022.

\bibitem{Dakkak2019AcceleratingRA}
Abdul Dakkak, Cheng Li, Isaac Gelado, Jinjun Xiong, and Wen mei W.~Hwu.
\newblock Accelerating reduction and scan using tensor core units.
\newblock {\em Proceedings of the ACM International Conference on
  Supercomputing}, 2019.

\bibitem{Dechev2006LockFreeDR}
Damian Dechev, Peter Pirkelbauer, and Bjarne Stroustrup.
\newblock Lock-free dynamically resizable arrays.
\newblock In {\em OPODIS}, 2006.

\bibitem{Feldman2016AnEW}
Steven~D. Feldman, Carlos Valera-Leon, and Damian Dechev.
\newblock An efficient wait-free vector.
\newblock {\em IEEE Transactions on Parallel and Distributed Systems},
  27:654--667, 2016.

\bibitem{Gelado2019ThroughputorientedGM}
Isaac Gelado and Michael Garland.
\newblock Throughput-oriented {GPU} memory allocation.
\newblock {\em Proceedings of the 24th Symposium on Principles and Practice of
  Parallel Programming}, 2019.

\bibitem{Hatipoglu1997ParallelTM}
Bilal Hatipoglu and Can~C. {\"O}zturan.
\newblock Parallel triangular mesh refinement by longest edge bisection.
\newblock {\em SIAM J. Sci. Comput.}, 37, 1997.

\bibitem{Huang2010XMallocAS}
Xiaohuang Huang, Christopher~I. Rodrigues, Stephen Jones, Ian Buck, and Wen mei
  W.~Hwu.
\newblock Xmalloc: A scalable lock-free dynamic memory allocator for many-core
  machines.
\newblock {\em 2010 10th IEEE International Conference on Computer and
  Information Technology}, pages 1134--1139, 2010.

\bibitem{Jenkins2018RCUArrayAR}
Louis Jenkins.
\newblock {RCUArray}: An {RCU}-like parallel-safe distributed resizable array.
\newblock {\em 2018 IEEE International Parallel and Distributed Processing
  Symposium Workshops (IPDPSW)}, pages 925--933, 2018.

\bibitem{King2016DynamicSA}
James King, Thomas Gilray, Robert~Michael Kirby, and Matthew Might.
\newblock Dynamic sparse-matrix allocation on gpus.
\newblock In {\em ISC}, 2016.

\bibitem{Mousa2021HighperformanceSO}
Mohamed-H. Mousa and M.~Hussein.
\newblock High-performance simplification of triangular surfaces using a gpu.
\newblock {\em PloS one}, 16 8:e0255832, 2021.

\bibitem{Navarro2014ASO}
Crist{\'o}bal~A. Navarro, Nancy Hitschfeld-Kahler, and Luis Mateu.
\newblock A survey on parallel computing and its applications in data-parallel
  problems using {GPU} architectures.
\newblock {\em Communications in Computational Physics}, 15:285--329, 2014.

\bibitem{8855701}
Can Peng, Chenlin Huang, Daokun Hu, Di~Bang, Jianhua Sun, Hao Chen, and Xionghu
  Zhong.
\newblock Address randomization for dynamic memory allocators on the gpu.
\newblock In {\em 2019 IEEE 21st International Conference on High Performance
  Computing and Communications; IEEE 17th International Conference on Smart
  City; IEEE 5th International Conference on Data Science and Systems
  (HPCC/SmartCity/DSS)}, pages 570--577, 2019.

\bibitem{Perry2020LowLevelMM}
Cory Perry and Nikolay Sakharnykh.
\newblock Introducing low-level gpu virtual memory management, 2020.

\bibitem{Sha2017AcceleratingDG}
Mo~Sha, Yuchen Li, Bingsheng He, and Kian-Lee Tan.
\newblock Accelerating dynamic graph analytics on gpus.
\newblock {\em Proc. VLDB Endow.}, 11:107--120, 2017.

\bibitem{Stotko2019stdgpuES}
Patrick Stotko.
\newblock stdgpu: Efficient stl-like data structures on the gpu.
\newblock {\em ArXiv}, abs/1908.05936, 2019.

\bibitem{Winter2020OuroborosVQ}
Martin Winter, Daniel Mlakar, Mathias Parger, and Markus Steinberger.
\newblock Ouroboros: virtualized queues for dynamic memory management on
  {GPUs}.
\newblock {\em Proceedings of the 34th ACM International Conference on
  Supercomputing}, 2020.

\bibitem{Winter2021AreDM}
Martin Winter, Mathias Parger, Daniel Mlakar, and Markus Steinberger.
\newblock Are dynamic memory managers on {GPUs} slow?: a survey and benchmarks.
\newblock {\em Proceedings of the 26th ACM SIGPLAN Symposium on Principles and
  Practice of Parallel Programming}, 2021.

\bibitem{Xiao2010InterblockGC}
Shucai Xiao and Wu~chun Feng.
\newblock Inter-block {GPU} communication via fast barrier synchronization.
\newblock {\em 2010 IEEE International Symposium on Parallel \& Distributed
  Processing (IPDPS)}, pages 1--12, 2010.

\end{thebibliography}

\end{document}